\documentclass[twocolumn, prl, aps, floats]{revtex4}

\DeclareMathAlphabet{\mathpzc}{OT1}{pzc}{m}{it}

\newcommand{\bftheta}{\boldsymbol{\theta}}
\newcommand{\bfn}{\boldsymbol{n}}
\newcommand{\bfN}{\boldsymbol{N}}

\makeatletter
\renewcommand{\fnum@figure}{Figure \thefigure}
\makeatother

\usepackage{amsmath, amssymb, bbm, bm, color, epsf, epsfig, float, graphicx, ragged2e, relsize, setspace, yfonts, xcolor}

\begin{document}

\begin{small}

\title{\flushleft
\textsf{\textbf{\huge The mathematical structure of innovation \vspace{6pt}}} \\
\textsf{\normalsize Thomas M. A. Fink$^*$ and Ali Teimouri$^\dagger$} \\
\textsf{\small London Institute for Mathematical Sciences, 35a South St, Mayfair, London W1K 2XF, UK}
\vspace*{-14pt}
}


\maketitle
\vspace{0pt}

\begin{justify}
\linespread{1}\footnotesize 
\noindent
Despite our familiarity with specific technologies, the origin of new technologies remains mysterious.
Are new technologies made from scratch, or are they built up recursively from new combinations of existing technologies?
To answer this, we introduce a simple model of recursive innovation in which technologies are made up of components and combinations of components can be turned into new components---a process we call technological recursion.
We derive a formula for the extent to which technological recursion increases or decreases the likelihood of making new technologies.
We test our predictions on historical data from three domains and find that technologies are not built up from scratch, but are the result of new combinations of existing technologies. 
This suggests a dynamical process by which known technologies were made and a strategy for accelerating the discovery of new ones.
\end{justify}
\vspace{0.08in}
\noindent 
Where do new technologies come from?
Despite our familiarity with specific technologies, such as wind turbines, touch screens and GPS,
how new technologies come into being remains mysterious \cite{Science}.
By definition, new technologies lie just beyond our imagination \cite{Felin},
and navigating an expanding space of possibilities is challenging \cite{Tria,Loreto}.
\\ \indent 
However, we have a better understanding of what technologies are.
Technologies are made up of component building blocks \cite{CompsE, FinkA, FinkNat, FinkAA, CompsC, CompsD}:
``a combination of components to some purpose" \cite{ArthurA}.
For example, GPS is made up of the components of atomic clocks, satellites and receivers,
and wind turbines are made up of rotors, generators and towers.
\\ \indent 
When a technology---a combination of components---becomes a reliable functioning unit, it becomes more readily available, and obtaining the whole is easier than gathering the individual parts.
It is given a name and used as a component in its own right.
So technologies are combinations of components, and these technologies can themselves be adopted as components.
In this way technologies are made up of more primitive technologies, in a hierarchical way, as shown in Fig.\ \ref{gps} for GPS.
New technologies are not built up from scratch, but are the result of new combinations of existing technologies---a hypothesis described qualitatively by Brian Arthur in \emph{The Nature of Technology} \cite{ArthurB}.
\\ \indent 
In the mind of the innovator, too, the combination is no longer a complex assembly, but a distinct unit, ready for use~\cite{McNerney}.
This alters the innovator's perception of the adjacent possible: 
it biases it in favour of designs which use the combination, since now they are one step away, rather than the multiple steps needed for the constituent parts.
\\ \indent 
\\ \noindent \textsf{\textbf{Technological recursion}} \\
Turning a combination of components into a new component is a recursive process, and we call it technological recursion.
How does it effect a technology sector?
On the one hand, technological recursion makes a sector simpler.
Technologies which were previously composed of many components become composed of fewer components.
This makes them easier to make because, given access to a fixed number of components, 
we are more likely to have the components needed to make technologies with few components than those with many \cite{FinkAA}.
\\ \indent 
On the other hand, technological recursion makes a technology sector more complicated. 
There is another component to keep track of, 
and the number of combinations of components grows exponentially with the number of components. 
In this expanded search space, there are more candidate combinations to sift through when searching for new technologies.
\\ \indent 
Too much technological recursion and too little both inhibit discovering new technologies.
With too much, there is a surge of new components, most of which are nearly useless.
This is similar to how, in language, if every sentence fragment becomes a new word, the numbers of words escalates.
With too little recursion, technologies become ever more complex, with no apparent structure across different organizational length scales.
For example, without technological recursion, GPS is made from at least 13 components rather than three (see Fig.\ \ref{gps}).
This is similar to how, in language, if no new words are introduced, the number of words needed to describe innovations  multiplies; 
without the new word ``GPS'', for instance, we would be stuck with ``global positioning system''.
\\ 
\\ \noindent \textsf{\textbf{In this Letter}} \\
The interplay between existing technologies and the components for making new ones 
is at the heart of understanding where new technologies come from.
In this Letter we do four things.
First, we introduce a simple model of recursive innovation in which technologies are made up of components, 
and combinations of components can be turned into new components---a process we call technological recursion.
Second, we derive a formula for the expected benefit of technological recursion: the extent to which it increases or decreases the likelihood of making new technologies.
Third, we apply our insights to historical data from language, gastronomy and drugs, and find that only a small fraction of technological recursions increase the chance of making new technologies.
Remarkably, these are almost always complete technologies, 
confirming the hypothesis that technologies are built up recursively. 
Fourth, we show that repeated technological recursion can increase the likelihood of making new technologies by orders of magnitude. 
This suggests a dynamical process by which known technologies were made,
and a strategy for accelerating the discovery of new ones~\cite{arthurpolak}.
\begin{figure}[b!]
\includegraphics[width=0.67\linewidth]{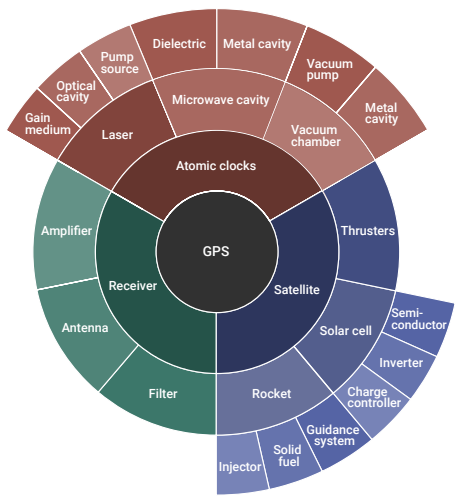}
\caption{
\sf\footnotesize \textbf{The recursion structure of technology.} 
Technologies are made up of more primitive technologies in a hierarchical way.
For example, GPS is made up of the components of atomic clocks, satellites and receivers.
Atomic clocks, in turn, are made up of lasers, microwave cavities and vacuum chambers; and so on.
We call this process of turning a combination of components into a new component technological recursion.
Although we do not show it here, GPS is itself a component in subsequent technologies, such as guided missiles and earthquake monitors.
}
\label{gps}
\end{figure}
\\ \\ \noindent \textcolor{red}{\textsf{\textbf{\large Results}}} 
\\ \noindent \textsf{\textbf{Lego game}} \\
To illustrate our insights, consider three children, Eve, Frank and Grace, playing with three different Lego sets.
The goal of each child is to make as many toys as possible.
Eve has an ordinary set of Lego bricks.
Frank has the same set, but he modified it by permanently gluing together pieces that frequently go together, such as wheels and axles, so that he has quick access to these modules when needed.
Grace adopted Frank's strategy, but took it further. 
Not only did she glue together pieces frequently used in the same toy, but she also glued pieces that are used in the same toy just occasionally, such as baseplates and doors.
Each child scoops out a random bunch of pieces from their box of Lego bricks. 
All three have the same number, but whereas Eve has only individual bricks, 
some of Frank's pieces are whole modules that he pre-assembled.
Grace has more modules and fewer individual bricks.
\\ \indent
Which child makes the most toys?
While Eve assembles individual bricks, 
Frank has the advantage that when the toy he is making contains one of his pre-assembled modules,
he's sure to have all the pieces contained in it.
But while Frank comes in first place, Grace comes in last.
Despite having more modules than Frank, her approach backfires. 
Many of her modules belong to toys that she can't complete.
Because she has fewer individual bricks than the others, she's often missing the right bricks to finish off toys she can partially make.
As we shall see, turning the right combinations of components into new components can dramatically increase the number of makeable technologies, but turning too many combinations into components has the opposite effect.
\\ \\ \noindent \textsf{\textbf{Components and technologies}} \\
We take technologies to be made up of distinct components \cite{ArthurA,ArthurB,FinkA,FinkAA,CompsE,CompsC,CompsD}.
A component can be a material object, like a capacitor, or a routine, like
polymerase chain reaction, or a skill, like coding in Java. 
Once a component has been discovered, we do not have to worry about running out; 
there are no capacity constraints.
Any subset of our components can be combined, but a combination either
is, or is not, a technology, according to some universal recipe book of technologies.
Suppose further that there are a total of $N$ possible components
in ``God's own cupboard''
but that, at any given stage $n$, we only have in our basket $n$ of these
$N$ possible components.
At every stage, we pick a new component to add to our basket, increasing $n$ by 1.
\\ \indent 
The size $c$ of a technology is the number of distinct components a technology is made of. 
The order of components is irrelevant, and multiple occurrences of a component count once, so the word ``innovation'' has $c = 6$ components, not 10. 
\\ \\ \noindent \textsf{\textbf{Size of the technology space}} \\
Consider a specific basket of $n$ components, which we call $\bfn$.
Notice how we differentiate between a particular set of components $\bfn$ and their number $n$.
Let the size of the technology space $p(\bfn)$ be the number of technologies that we can make from $\bfn$.
Of course, $p(\bfn)$ depends on the specific components in our basket, and not just on how many we have;
for example, we can make more words from the first five letters of the alphabet than the last five.
To bypass this, we take the average over all possible baskets of size $n$ drawn from the $N$ possible components.
We call this the expected size of the technology space, $\overline{p}(n)$.
We denote the number of makeable technologies of size $c$ by $p(\bfn,c)$, so that summing $p(\bfn, c)$ over $c$ gives $p(\bfn)$.
\\ \indent 
In previous work \cite{FinkAA,FinkB}, we proved that $\overline{p}(n,c)$---the average of $p(\bfn, c)$ over all possible baskets $\bfn$---satisfies an exact conservation law: for two different stages $n$ and $n'$,
\begin{equation}
\textstyle \overline{p}(n,c) \big/ \binom{n}{c} = \overline{p}(n',c) \big/ \binom{n'}{c},
\label{pConservationBody}
\end{equation}
where $\binom{n}{c}$ is the binomial coefficient.
We will use this later when we determine the effect of technological recursion.
\begin{figure}[b]
\centering
\includegraphics[width=0.86\linewidth]{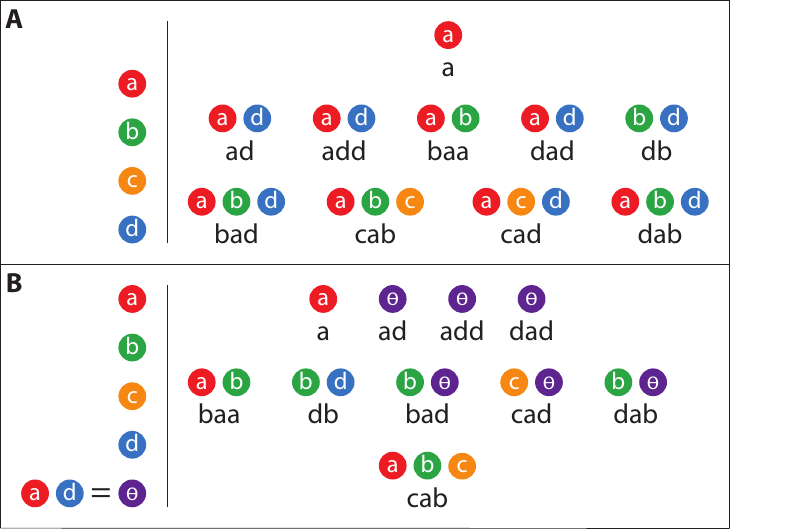}
\includegraphics[width=0.874\linewidth]{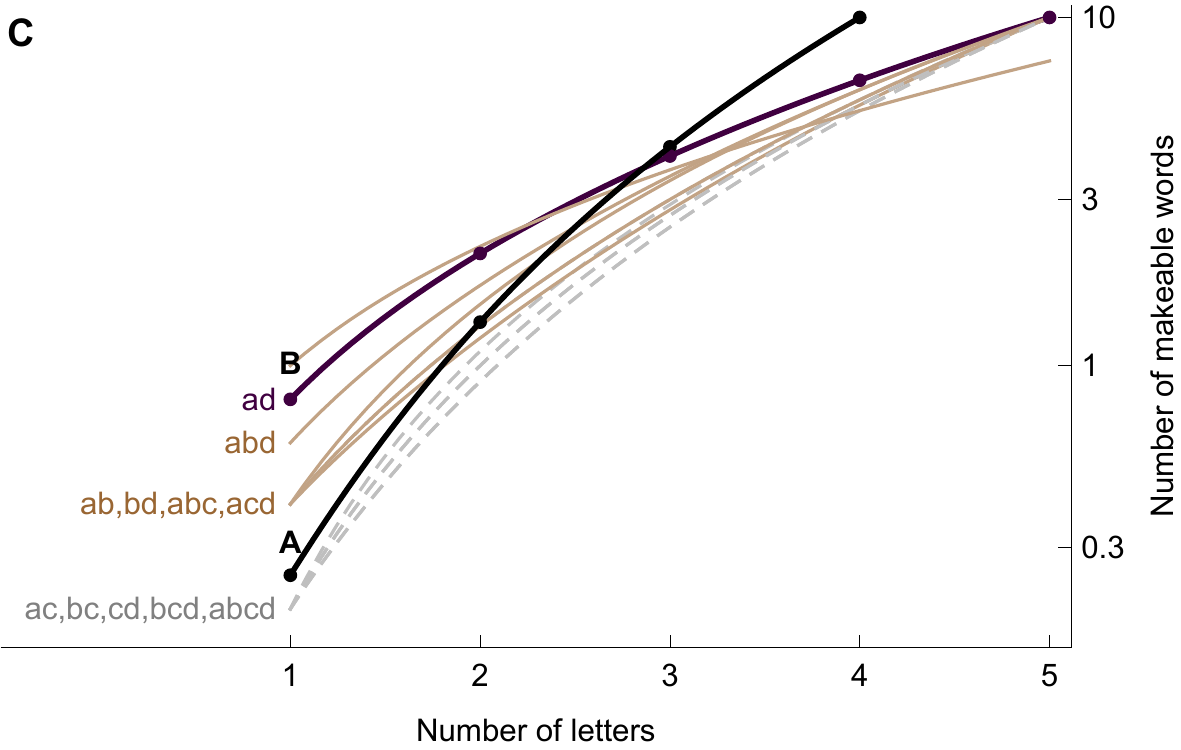}
\caption{\sf\footnotesize \textbf{Turning a technology into a component can increase the number of makeable technologies.} \textbf{A} In a simple example from language, the components are the letters a, b, c and d and the technologies are the 10 English words that can be made from them. 
The order and repetition of letters in the words are irrelevant.
One word has 1 different kind of component, five have 2 kinds and four have 3 kinds.
\textbf{B}
When we expand our repertoire of components by turning one of the words, ad, into a component---call it $\theta$---some words are made up of fewer components than before.
Now four words have 1 different kind of component, five have 2 kinds and one has 3 kinds. 
\textbf{C}
How many words can we make, on average, when we pick \emph{n} random components?
When we draw from the original repertoire of four components in \textbf{A}, the result is given by the black curve.
When we draw from the new repertoire of five components in \textbf{B}, it is given by the purple curve.
Turning ad into a component increases the expected number of makeable words 
3.2-fold when we draw one component and
1.6-fold when we draw two,
but when we draw three or more components it decreases.
The solid tan curves show the result for turning other words into components, all of which offer some advantage when the size of the draw is small.
The dashed gray curves show the result for all other combinations, none of which offer any advantage.
}
\label{FigToy}
\end{figure}
\begin{figure*}[t!]
\includegraphics[width=1\linewidth]{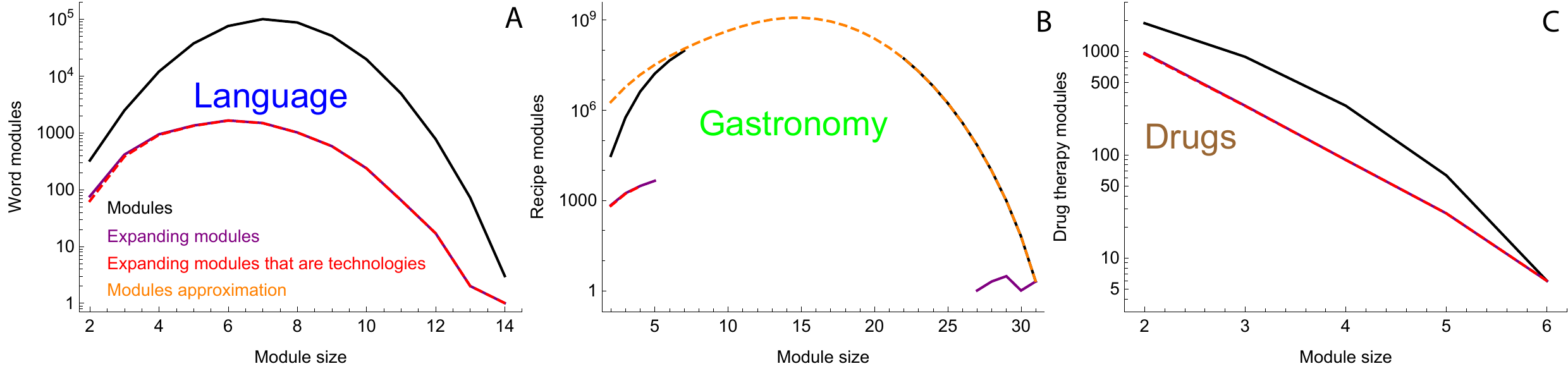} 
\caption{\sf\footnotesize 
\textbf{Expanding modules are rare but almost always complete technologies.} 
{\bf ABC} 
In our three domains, the top curves are the number of distinct combinations of $c$ components: pairs, triples, quadruples, and so on. 
The purple solid curve is the number expanding modules: combinations that increase the likelihood of making new technologies.
The red dashed curves are the number of expanding modules that are complete technologies, rather than parts of technologies.
Remarkably, almost all expanding modules are complete technologies. 
}
\label{fig:imageZ}
\end{figure*}
\\ \\ \noindent \textsf{\textbf{Combination usefulness}} \\
Some combinations of components show up in more technologies than others.
Consider some combination of $k$ components which appear together in one or more technologies;
we call the $k$ components a module and label it $\bftheta$ for convenience.
The combination usefulness $u_{\bftheta}(\boldsymbol {n})$ is the number of technologies makeable from the components $\boldsymbol{n}$ that the combination $\bftheta$ shows up in.
As with the size of the technology space $p$, $u_{\bftheta}(\bfn,c)$ denotes the number of technologies of size $c$ that $\bftheta$ shows up in.
\\ \indent 
We proved (see Methods) that $\overline{u}_{\bftheta}(n,c)$---the average of $u_{\bftheta}(\bfn, c)$ over all possible baskets $\bfn$---satisfies an exact conservation law: for two different stages $n$ and $n'$,
\begin{equation}
    \textstyle
    \overline{u}_{\bftheta}(n',c) \, \binom{n'}{k}\big/\binom{n'}{c} = {\overline{u}_{\bftheta}(n,c)} \, \binom{n}{k}\big/\binom{n}{c}.
    \label{uConservationBody}
\end{equation}
With these two conservation laws in hand, we are ready to determine the effect of technological recursion.
\\ \\ \noindent \textsf{\textbf{Effect of technological recursion}} \\
The likelihood that a random combination of $c$ components is a technology is $\overline{p}(n,c)/\binom{n}{c}$.
Therefore increasing the likelihood of making a technology corresponds to increasing the expected number of technologies $\overline{p}(n,c)$ that we can make, which we now address.
\\ \indent 
In a technological recursion,
we replace a combination of $k$ components $\bftheta$ with a new individual component $\theta$ in all of the technologies in which the combination $\bftheta$ shows up.
We indicate this as $\bftheta \rightarrow \theta$.
Notice that we differentiate between the combination of components $\boldsymbol{\theta}$ and the new individual component $\theta$ that replaces them.
An example is shown in Fig.\ \ref{FigToy}{\bf ab}.
\\ \indent 
When we apply the technological recursion $\bftheta \rightarrow \theta$, some technologies of size $c+k-1$ are reduced to size $c$, and some technologies of size $c$ are reduced to size $c-k+1$. 
So we both add and subtract to the original number of makeable technologies of size $c$.
The effect of a technological recursion depends on whether this net change, summed over $c$, is positive or negative.
We obtained a formula, derived in the Methods, that gives the expected number of technologies we can make after applying $\bftheta \rightarrow \theta$:
\begin{eqnarray}
\overline{p}_{\bftheta}(n)  \!&\simeq&\! \Big(1 - \textstyle \frac{x}{N+1} \frac{\rm d}{{\rm d}x} \Big) \big(\overline{p}(n) + (x - x^k) \, \overline{u}_{\bftheta}(n)   \big),
\label{pthetaBody}
\end{eqnarray}
where $x = n/N$.
Because $p(n)$ and $u_{\bftheta}(n)$ are unbiased estimates of their means, we can estimate
$\overline{p}(n)$ and $\overline{u}_{\bftheta}(n)$ by $p(\bfn)$ and $u_{\bftheta}(\bfn)$.
We do not assume knowledge of the technology recipe book for components that we do not possess; there is no omniscience.
\begin{figure*}[t!]
\includegraphics[width=1\linewidth]{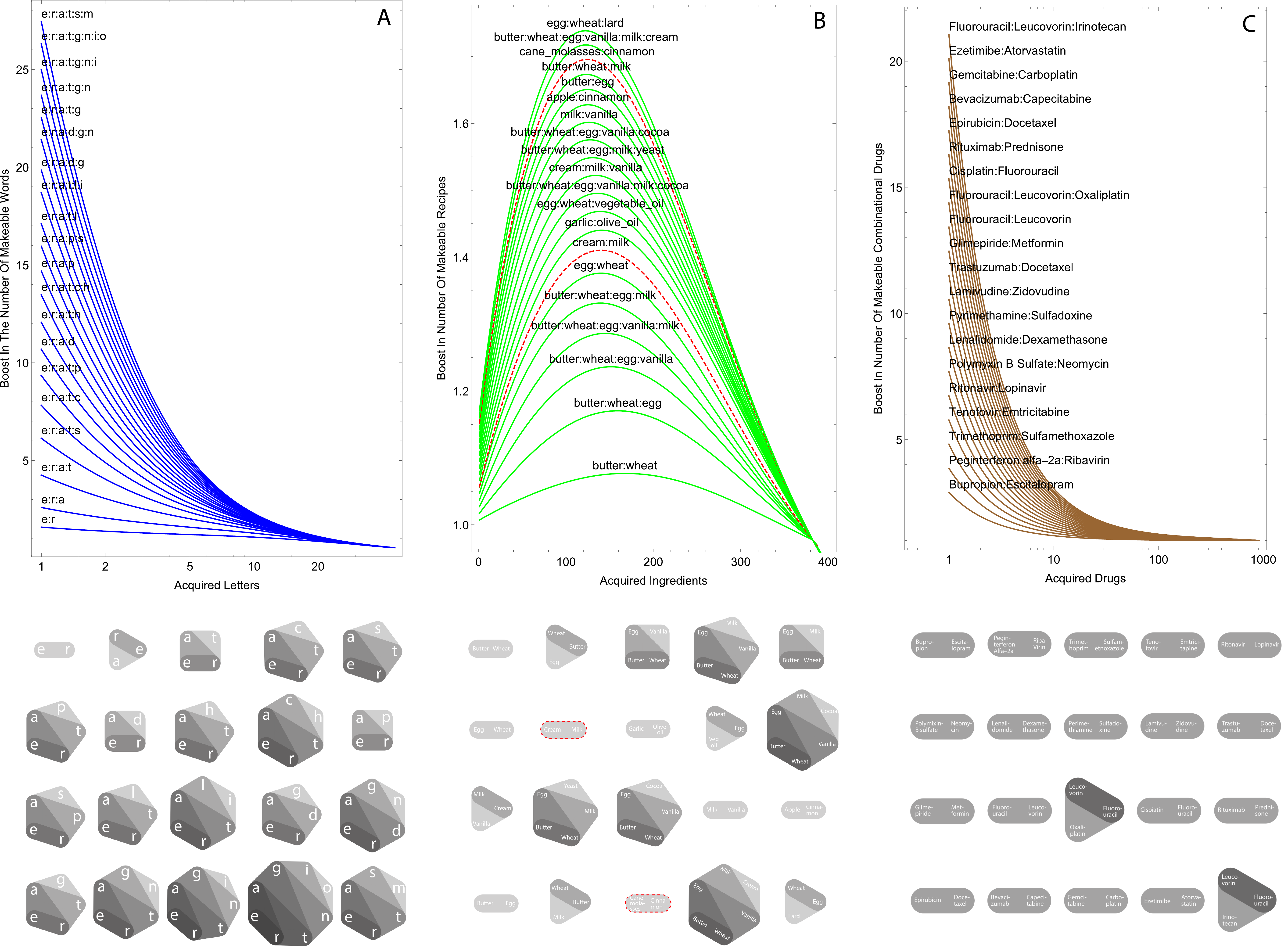} 
\caption{\sf\footnotesize 
\textbf{Recursively turning technologies into components. ABC}
For each domain, we applied a technological recursion using the best pair module (a module of size $k=2$). 
We repeated this process, recursively, 20 times. 
When a pair of components is subsequently used as a component in another pair, it appears as a triple of components, and so on.
The hierarchical structure of the 20 technological recursions is shown below each sector.
}
\label{fig:image2}
\end{figure*}
\\ \\ \noindent \textsf{\textbf{Expanding modules}} \\
When $\overline{p}_{\boldsymbol{\theta}}(n) > \overline{p}(n)$,
we can make more technologies from $n$ components drawn from the updated repertoire of $N+1$ components
than we can from $n$ components drawn from the original repertoire of $N$ components.
We call a combination $\bftheta$ that increases the number of makeable technologies an expanding module.
From (\ref{pthetaBody}), $\overline{p}_{\boldsymbol{\theta}}(n) > \overline{p}(n)$ implies
\begin{equation}
\textstyle
\overline{v}_{\bftheta}(n)  > \frac{x}{N+1} \big(\overline{p}'(n) + \overline{v}'_{\bftheta}(n)\big),
\end{equation}
where $x = n/N$, $\overline{v}_{\bftheta}(n) = (x- x^k) \, \overline{u}_{\bftheta}(n)$ and $'$ indicates the derivative with respect to $x$. 
\\ \\ \noindent \textsf{\textbf{Language, gastronomy and drugs}} \\
With a test for expanding modules in hand, we now apply our theory to real systems.
We gathered data from three domains: language, gastronomy and drugs.
In language,    the technologies are the 38,970 English words   and the components are the 26 letters used to make them.
In gastronomy,  the technologies are 56,498 recipes             and the components are the 381 ingredients used to make them  \cite{ahnert14}.
In drugs,       the technologies are the 1363 known drug combination therapies and the components are the 901 individual drugs used to make them \cite{Liu}.   
\\ \indent 
We did the following experiment for each domain.
We randomly selected $n$ components to put into our basket and counted how many technologies we could make from them.
Using the conservation law in eq. (\ref{pConservationBody}), we were able to average this over all possible baskets of size $n$.
This gives $\overline{p}(n)$.
\\ \indent 
We then picked a combination of $k$ components $\bftheta$,
and replaced it with the new individual component $\theta$ everywhere it showed up in a technology.
This raises the total number of possible components from $N$ to $N+1$. 
We randomly selected $n$ components drawn from this new repertoire to put into our basket and counted how many technologies we could make.
As before, we averaged this over all possible baskets using eq. (\ref{pthetaBody}).
This gives $\overline{p}_{\bftheta}(n)$.
\\ \\ \textsf{\textbf{Expanding combinations are rare}} \\
We repeated the above experiment for all possible combinations of $k$ components $\bftheta$ that show up in at least one technology: pairs, triples, quadruples and so on.
(We did this for all $k$ in language and drugs but for $k\leq 5$ for gastronomy for computational reasons.)
We found that only a small fraction of these combinations increase the number of makeable technologies.
For $k\leq 5$ across all domains, we found that
in language,    there are 128,413       letter combinations,        of which 4,426 are expanding;
in gastronomy,  there are 65,157,341    ingredient combinations,    of which 9,839 are expanding; and
in drugs,       there are 3,130        drug combinations,          of which 1,381 are expanding.
These are plotted in  Fig.\ \ref{fig:imageZ} as a function of the module size $k$.
\\ \\ \textsf{\textbf{Expanding modules are complete technologies}} \\
Our hypothesis is that technologies are not built up from scratch, but are the result of new combinations of existing technologies \cite{ArthurA,ArthurB}. 
In other words, the components of technologies are themselves complete technologies, and not just parts thereof.
\\ \indent 
To test this this hypothesis, we checked to see how many of these expanding modules are complete technologies, and not just parts of technologies.
Remarkably, almost all expanding modules are complete technologies: 98.70\%, 96.88\% and 98.62\% in language, gastronomy and drugs.
\\ \\ \noindent \textsf{\textbf{Recursively turning technologies into components}} \\
So far we have applied just a single technological recursion to our domains.
But we can repeatedly turn technologies into components in a recursive way.
In this case the effects combine to amplify each other. 
\\ \indent 
For each of our three domains, we found the most expanding pair module (a module of size $k=2$) and applied it as a technological recursion.
We then repeated the process, recursively, always only considering pair modules.
The result of 20 iterations of this recursive process are shown in Fig.\ \ref{fig:image2}, top. 
We find that sometimes, the best pair module involves previously formed pairs, shown in Fig.\ \ref{fig:image2}, bottom.
The end result is that the number of makeable technologies increases by a factor of two to 28.
\\ \\ \noindent \textcolor{red}{\textsf{\textbf{\large Discussion}}} 
\\ \textsf{\textbf{Modularity and technological recursion}} \\
In any collection of systems of parts, certain motifs may recur: groups of parts which regularly appear together or relate to each other in the same way.
A system exhibits modularity when the recognition of these motifs makes it possible to describe it more concisely or manipulate it more easily \cite{Clark,HidalgoB,LucianoB}.
One of the challenges in identifying modules is that there is, in general, no objective criterion for when a motif is sufficiently regular to be recognized and treated as a unit in its own right.
Just as data analysis is fraught with the possibility of under-fitting and over-fitting, a system of many parts can be under-modularized and over-modularized.
At the extremes, every repeating motif is a module and no motif is sufficiently common to be a module.
\\ \indent 
We provide an objective criterion for identifying modules in technology, namely, that a combination of components forms a module if doing so increases the likelihood of discovering new technologies.
We find that such combinations of components are rare,
and that they have a recursive structure: they possess a hierarchy of organizational length scales which increases as new technologies are discovered~\cite{Langlois}.
But that's not all.
We also find that the modules present in technologies are nearly always themselves complete technologies, 
rather than parts of technologies or a collection of disparate components from multiple technologies. 
This provides quantitative support for the hypothesis that technologies are the result of new combinations of existing technologies \cite{ArthurA, ArthurB}.
\\ \noindent 
\\ \textsf{\textbf{Transforming the adjacent possible}} \\
Technological innovation can be conceived of as exploring an expanding space of possibilities.
Each new building block combines with our existing set of building blocks in new ways.
From this perspective, technological recursion---turning a combination of components into a new component---transforms the space by bringing regions that were once far apart closer together.
When a technology of multiple components becomes a single component, it is easier to use that technology to make a new technology because using it is now one step away, 
rather than the multiples steps required to reach all of its constituent components.
In this way technological recursion alters the map of the adjacent possible \cite{Tria,Loreto}. 
The innovator, especially one with limited resources, is biased towards technologies nearby in possibility space because there are fewer parts that he might be lacking \cite{PRL2}.
Our insights provide a prescription for which technological recursions maximise the likelihood of discovering new technologies \cite{Emerging1}. 
\\ \noindent 
\\ \textsf{\textbf{New technologies shape their own environment}} \\
In a recursive process, the output of the process becomes the input to the same process.
Recursive processes are hard to understand because they can shape their own environment;
we can no longer assume that the environment is constant, or at least slowly changing, as is commonly assumed in models of evolution, for example.
In technological recursion, the process is combining components, and the output---a potential technology---can become part of the input---the repertoire of components for making new technologies.
\\ \indent 
An important consequence of technological recursion is that technologies shape their own environment:
the likelihood of new technologies coming into being can be substantially altered by the the technology itself.
While the technologies of next year will be made up in part of the technologies of today, 
the technologies two years hence will be made up of those of next year. 
Forecasting technology thus involves a hierarchy of prediction problems in an expanding space of building blocks,
which helps explain the poor track record of long-term technological prediction \cite{Felin,FarmerB}.
\\ \\ \noindent \textsf{\textbf{Drug discovery}} \\
Drug combination therapies, one of the three domains that we analyzed, use multiple drugs simultaneously to treat a disease. 
But designing new combination therapies is difficult because there are a thousand individual drugs which can be used as components \cite{Liu}. 
So there are on the order of a million possible two-drug therapies, a billion three-drug therapies, and so on. 
While there are some rules of thumb for which drugs tend to co-appear in successful therapies, thereby reducing the space of possibilities, drug combination therapy lacks systematic design principles \cite{Liu}.
Similar challenges are found in combinatorial chemistry, which has played a key role in the discovery of many drugs. 
But such brute force approaches, with their low odds of success, are expensive,
and the way in which many pharmaceutical companies approach drug discovery is thought to be unsustainable \cite{Hunter}.
\\ \indent 
Using our insights, pharmaceutical companies can increase the chance of hitting upon new drugs and drug combination therapies by strategically redefining the repertoire of component building blocks to include existing drugs and drug therapies~\cite{dennisverhoeven,Strumsky}. 
This is analogous to how, in language, phonemes can be a more successful repertoire of components for building words than letters.
Our analysis could be enhanced by access to the proprietary positive and negative results held by drug firms.
This would lead to higher discovery hit rates and fewer wasted trials.
\\ \\ \noindent \textcolor{red}{\textsf{\textbf{\large Methods}}} 
\\ \noindent \textsf{\textbf{Data}} \\
Our three data sets were obtained as follows. In language, our list of common English words is from the built-in WordList library in Mathematica 11.3. 
Of the 39,176 words in WordList, we only considered the 38,970 made from
the 26 letters a--z, ignoring case: we excluded words containing a hyphen,
space, and so on. 
In gastronomy, the 56,498 recipes can be found in the supplementary material in \cite{ahnert14}.
In drugs, the 1,363 therapies can be found in \cite{Liu}.
\\ \\ \noindent \textsf{\textbf{Module invariant}} \\
Let $\boldsymbol{N}$ be the set of $N$ possible components and $\bftheta$ be a subset of $k$ components, or module, chosen from $\boldsymbol{N}$. 
Let $\boldsymbol{N}_k$ be the set of $N-k$ other components not including $\bftheta$,
let $\boldsymbol{n}_k$  be a subset of $n-k$ components chosen from $\boldsymbol{N}_k$, and
let $\boldsymbol{c}_k$ be a subset of $c-k$ components chosen from $\boldsymbol{n}_k$.
The module usefulness $u_{\bftheta}(\boldsymbol {n},c)$ is the number of technologies of size $c$ makeable from the components $\boldsymbol{n}$ that the $k$ components $\bftheta$ show up in.
We can also think of the module usefulness in a different way. 
It is how many more technologies of size $c$ we can make from the components $\boldsymbol{n}_k$ together with $\bftheta$, 
than from the components $\boldsymbol{n}_k$ alone:
\begin{equation*}
u_{\bftheta}(\boldsymbol {n},c) = \sum_{\boldsymbol{c}_k \subseteq \boldsymbol{n}_k} \mathcal{P}(\bftheta \cup \boldsymbol{c}_k)  - \mathcal{P}(\boldsymbol{c}_k),
\end{equation*}
where $\mathcal{P}(\bftheta \cup \boldsymbol{c}_k)$ takes the value 0 if the combination of components $\bftheta \cup \boldsymbol{c}_k$ form no technologies of size $c$ and 1 if $\theta \cup \boldsymbol{c}_k$ form one technology of size $c$.
\\ \indent
The mean usefulness of the $k$ components $\bftheta$, $\overline{u}_{\bftheta}(n,c)$, is the average of $u_{\theta}(\boldsymbol{n},c)$ over all subsets $\boldsymbol{n}_k \subseteq \boldsymbol{N}_k$;
there are $\binom{N-k}{n-k}$ such subsets.
Therefore
\begin{eqnarray*}
\overline{u}_{\bftheta}(n,c)        &=& {1}\big/{\textstyle{N-k \choose n-k}}  \sum_{\boldsymbol{n}_k \subseteq \boldsymbol{N}_k} u_{\bftheta}(\bfn,c) \label{PP} \\
                                        &=& {1}\big/{\textstyle{N-k \choose n-k}}  \sum_{\boldsymbol{n}_k \subseteq \boldsymbol{N}_k}  \,\, \sum_{\boldsymbol{c}_k \subseteq \boldsymbol{n}_k}  \mathcal{P}(\bftheta \cup \boldsymbol{c}_k)  - \mathcal{P}(\boldsymbol{c}_k).
\end{eqnarray*}
Consider some particular combination of components $\boldsymbol{c}_k'$.
The double sum above will count $\boldsymbol{c}_k'$ once if $c=n$, but multiple times if $c < n$, 
because $\boldsymbol{c}_k'$ will belong to multiple sets $\boldsymbol{n}_k$.  
How many?
In any set $\boldsymbol{n}_k$ that contains $\boldsymbol{c}_k$, there are $n-c$ free elements to choose, from $N-c$ other components.
Therefore the double sum will count every combination $\boldsymbol{c}_k$ a total of ${N-c \choose n-c}$ times, and
\begin{eqnarray}
\overline{u}_{\bftheta}(n,c)    &=&                                             {\textstyle {N-c \choose n-c}} \big/ {\textstyle{N-k \choose n-k}}
                                                                                                \sum_{\boldsymbol{c}_k \subseteq \boldsymbol{N}_k} \mathcal{P}(\bftheta \cup \boldsymbol{c}_k)  - \mathcal{P}(\boldsymbol{c_k}) \nonumber \\
                                &=& \textstyle \binom{n}{c}\big{/}\binom{N}{c} \, \binom{N}{k}\big{/}\binom{n}{k}  \, u_{\bftheta}(\bfN,c),
\label{unNMethods}
\end{eqnarray}
noting that $\overline{u}_{\bftheta}(N,c) = u_{\bftheta}(\bfN,c)$.
Solving both equations for $\overline{u}_{\bftheta}(N,c)$ and equating them, we obtain the exact conservation law for the module usefulness for any two stages $n'$ and $n$:
\begin{equation*}
    \textstyle
    \overline{u}_{\bftheta}(n'\!,c) \, \binom{n'}{k}\big{/}\binom{n'}{c} = {\overline{u}_{\bftheta}(n,c)} \, \binom{n}{k}\big{/}\binom{n}{c}.
    \label{exact}
\end{equation*}
\noindent \textsf{\textbf{Technological recursion equation}} \\
Setting $n' = N$ in eq.\ (\ref{pConservationBody}), $\overline{p}(n,c)$ can be expressed in terms of the number of makeable technologies when we have access to all $N$ of the possible components,
\begin{equation}
        \textstyle
        \overline{p}(n,c) = p(\boldsymbol{N},c) \binom{n}{c}\big{/}\binom{N}{c},
        \label{pnNMethods}
\end{equation}
noting that $\overline{p}(N,c) = p(\bfN,c)$.
\\ \indent 
Now let us apply the technological recursion $\bftheta \rightarrow \theta$, that is, let us replace the combination of $k$ components $\bftheta$ with the single new component $\theta$ every time the $k$ components show up in the same technology.
Let $\overline{p}_{\bftheta}(n,c)$ be the expected number of technologies of size $c$ that can be made from $n$ components drawn from $\boldsymbol{N}\cup \theta$, the union of the $N$ components $\boldsymbol{N}$ and the new component $\theta$, giving $N+1$ components in total.
Just as in eq.\ (\ref{pnNMethods}), we can write
\begin{equation}
    \overline{p}_{\bftheta}(n,c) = \textstyle p(\boldsymbol{N}\cup \theta,c) {n \choose c} \big/ {N+1 \choose c}.
    \label{qinitial}
\end{equation}
\indent
When we replace $\bftheta$ with $\theta$, some technologies of size $c+k-1$ are reduced to size $c$, and some technologies of size $c$ are reduced to size $c-k+1$. 
It is the net change that counts, so the relation between $p(\boldsymbol{N} \cup \theta,c)$ and $p(\boldsymbol{N},c)$ is
\begin{equation*}
        p(\bfN \cup \theta,c) = p(\boldsymbol{N}\!,c) 
        - u_{\bftheta}(\bfN\!,c)
        + u_{\bftheta}(\bfN\!,c+k-1).
\end{equation*}
Substituting this into eq.\ (\ref{qinitial}) gives
\begin{eqnarray*}
\overline{p}_{\bftheta}(n,c)    &=&  \big(p(\bfN\!,c) -u_{\bftheta}(\bfN\!,c) + u_{\bftheta}(\boldsymbol{N}\!,c+k-1)\big) \\
                                &\times& \Big(1 - {\textstyle \frac{c}{N+1}}\Big) \, \textstyle \binom{n}{c}/\binom{N}{c}.
\end{eqnarray*}
Substituting eqs. (\ref{pnNMethods}) and (\ref{unNMethods}) into this, 
and approximating $\binom{n}{c}$ and $\binom{N}{c}$ by $n^c$ and $N^c$ for $n,N \gg c$, we find
\begin{eqnarray*}
\overline{p}_{\bftheta}(n,c)  \!\!&\simeq&\!\!  \Big(1 \!-\! {\textstyle \frac{c}{N+1}}\Big) 
                                        \big(\overline{p}(n,c) \!-\! x^k \, \overline{u}_{\bftheta}(n,c) \!+\! x \, \overline{u}_{\bftheta}(n,c+k-1) \big)\\
                            \!\!&=&\!\!  \Big(1 \!-\! \textstyle \frac{x}{N+1} \frac{\rm d}{{\rm d}x} \Big) \big(\overline{p}(n,c) \!-\! x^k \, \overline{u}_{\bftheta}(n,c) 
\!+\! x \, \overline{u}_\theta(n,c\!+\!k\!-\!1) \big),
\end{eqnarray*}
where $x = n/N$.
Summing over $c$,
\begin{eqnarray}
\overline{p}_{\bftheta}(n)  \!&\simeq&\! \Big(1 - \textstyle \frac{x}{N+1} \frac{\rm d}{{\rm d}x} \Big) \big(\overline{p}(n) + (x- x^k) \, \overline{u}_{\bftheta}(n) \big).
\label{qmaster}
\end{eqnarray}
\noindent \textsf{\textbf{Toy model}} \\
Despite its simplicity, the toy model in Fig.\ \ref{gps} captures many of our insights, and it is worth working through it explicitly.
From eq.\ (\ref{pnNMethods}), the expected number of makeable words when we randomly draw $1 \leq n \leq 4$ letters from a, b, c and d is
\begin{eqnarray*}
\overline{p}(n)  &=& \textstyle 1 {n \choose 1} / {4 \choose 1} + 5  {n \choose 2} / {4 \choose 2} + 4 {n \choose 3} / {4 \choose 3} \\
                            &=& (2n^3 - n^2 + 2n)/12,
\end{eqnarray*}
where the coefficients 1, 5 and 4 are the numbers of words of size 1, 2 and 3.
\\ \indent
Now let us replace the letters a and d---which together we call $\bftheta$---with a new single component $\theta$ whenever a and d show up in the same word.
The expected number of makeable words when we randomly draw $n \in [1,5]$ letters from a, b, c and d and $\theta$ is
\begin{eqnarray*}
\overline{p}_{\bftheta}(n)   &=&     \textstyle         4 {n \choose 1} / {5 \choose 1} + 5  {n \choose 2} / {5 \choose 2} + 1 {n \choose 3} / {5 \choose 3} \\
                    &=&     (n^3 + 12n^2 + 35n)/60,
\end{eqnarray*}
where the coefficients 4, 5 and 1 are the numbers of words of size 1, 2 and 3 after the technological recursion $\bftheta \rightarrow \theta$.
For the sake of visualization, in Fig.\ \ref{gps} we interpolate between integer values of $n$ by replacing the factorials in the binomial coefficients  with the gamma function, even though such non-integer values are unphysical.
\\ \indent
We can calculate $\overline{p}_{\bftheta}(n)$ without explicitly making the substitutions by using eq.\ (\ref{qmaster}). 
We find
\begin{eqnarray*}
\overline{p}_{\bftheta}(n)    &=& \textstyle \frac{4}{5}        (1+3-0) {n \choose 1} / {4 \choose 1} +\textstyle \frac{3}{5}      (5+3-3) {n \choose 2} / {4 \choose 2} + \\
                        && \textstyle \frac{2}{5}       (4+0-3) {n \choose 3} / {4 \choose 3} \\
                        &=& (n^3 + 12n^2 + 35n)/60,
\end{eqnarray*}
which matches the result of explicit substitution above.
\\ \indent
\\ \\ \noindent \textsf{\textbf{Approximating the number of modules}} \\
We calculate the number of modules by counting the subsets of size $k$ of each technology (pairs, triples, and so on)
and getting rid of duplicates. 
For data sets that have technologies with many components, such as gastronomy where the largest recipes has 31 ingredients, this is computationally demanding. 
In this case we approximate the number of modules by assuming that the likelihood a subset of one technology belongs to another technology is negligible, the number of modules is then given by
\begin{equation}
\sum_{c\geq k} p(N,c) \textstyle \binom{c}{k}.
\end{equation}
This is the dashed orange line in Fig.\ \ref{fig:imageZ}, which matches the true number well for $c \geq 6$.
\begin{justify}
\linespread{1}\scriptsize\sf
\noindent \textbf{Acknowledgements} \\
The authors thank Martin Reeves and Andriy Fedosyeyev for helpful discussions and Roman Rybiansky for assisting with the figures.
A. Teimouri's postdoc position is funded by the BCG Henderson Institute. 
\\ \\
\noindent
$^*$tf@lims.ac.uk, $^\dagger$it@lims.ac.uk
\end{justify}

\end{small}
\end{document}